\documentclass[10pt,twocolumn,twoside,english]{IEEEtran}
\usepackage[T1]{fontenc}
\usepackage{mathrsfs}
\usepackage{bm}
\usepackage{amsmath}
\usepackage{amsthm}
\usepackage{amssymb}
\usepackage{algorithmic}
\usepackage[ruled,vlined,linesnumbered]{algorithm2e}
\usepackage{graphicx}
\usepackage{adjustbox}
 \PassOptionsToPackage{normalem}{ulem}

\makeatletter

\theoremstyle{plain}

  \theoremstyle{plain}
  
  \theoremstyle{plain}
  
  \theoremstyle{plain}
   
  \theoremstyle{remark}

\theoremstyle{assumption}
    \theoremstyle{proposition}

\theoremstyle{algorithm}  
   \theoremstyle{plain}
   
\usepackage{amsmath}
\usepackage{amssymb}
\usepackage{graphicx,psfrag,cite,subfigure}
\usepackage[table]{xcolor}

\setkeys{Gin}{width=1.0\columnwidth}

\makeatother

\usepackage{babel}
  \providecommand{\definitionname}{Definition}
  \providecommand{\lemmaname}{Lemma}
  \providecommand{\propositionname}{Proposition}
  \providecommand{\remarkname}{Remark}
\providecommand{\theoremname}{Theorem}
\providecommand{\conjecturename}{Conjecture}
\providecommand{\assumptionname}{Assumption}

\begin{document}

 \title{Efficient detection of adversarial images 
 \thanks{The authors are with the Department of Electrical Engineering, Indian Institute of Technology (IIT), Delhi. Arpan Chattopadhyay is also associated with the Bharti School of Telecom  Technology and Management, IIT Delhi. Email:  \{ee3180534, ee3160495, ee3160503, arpanc, ink\}@ee.iitd.ac.in }
 \thanks{This work was supported by the faculty seed grant, professional development allowance (PDA) and professional development fund (PDF) of Arpan Chattopadhyay at IIT Delhi.
}
\thanks{This manuscript is an extended version of our conference paper \cite{mundra2020adversarial}}
}

\author{
Darpan Kumar Yadav, Kartik Mundra, Rahul Modpur, Arpan Chattopadhyay, Indra Narayan Kar
}

\maketitle

%
%



\ifdefined\SINGLECOLUMN
	\setkeys{Gin}{width=0.5\columnwidth}
	\newcommand{\figfontsize}{\footnotesize} 
\else
	\setkeys{Gin}{width=1.0\columnwidth}
	\newcommand{\figfontsize}{\normalsize} 
\fi

\begin{abstract}
  In this paper, detection of deception attack on deep neural network (DNN) based image classification in autonomous and cyber-physical systems is considered. Several studies have shown the vulnerability of  DNN to malicious deception attacks. In such attacks, some or all pixel values  of an image are modified by an external attacker, so that the change is almost invisible to human eye but significant enough  for a DNN-based classifier to misclassify it. This paper first  proposes a novel pre-processing technique that facilitates detection of such modified images under any DNN-based image classifier as well as attacker model.  The proposed pre-processing algorithm involves a certain combination of principal component analysis (PCA)-based decomposition of the image, and random perturbation based detection  to reduce computational complexity. Next, an adaptive version of this algorithm is proposed where a random number  of perturbations are chosen adaptively using a doubly-threshold policy, and the threshold values are learnt via stochastic approximation in order to minimize the expected number of perturbations subject to constraints on the false alarm and missed detection probabilities.   Numerical experiments show that the proposed detection scheme  outperform a competing   algorithm while achieving  reasonably low  computational complexity. 
\end{abstract}


   
\maketitle

\section{Introduction}
Recently there have been significant research interest in
 cyber-physical systems (CPS) that connect the
cyber world and the physical world, via  integration of sensing, control, communication, computation  and learning. 
Popular CPS applications include networked monitoring of industry, disaster management, smart grids, intelligent transportation systems, networked surveillance, etc. One important component of future intelligent transportation systems is autonomous vehicle. It is envisioned that future autonomous vehicles will be equipped with high-quality cameras, whose images will be classified by a DNN-based classifier for object detection and recognition, in order to facilitate an informed maneuvering decision by the controller or autopilot. Clearly, vehicular safety in such cases is highly sensitive to  image classification; any mistake in object detection or classification can lead to accidents. In the context of surveillance or security systems,  adversarial  images can greatly endanger human and system security.

\begin{figure*}
  \includegraphics[width=0.8\paperwidth]{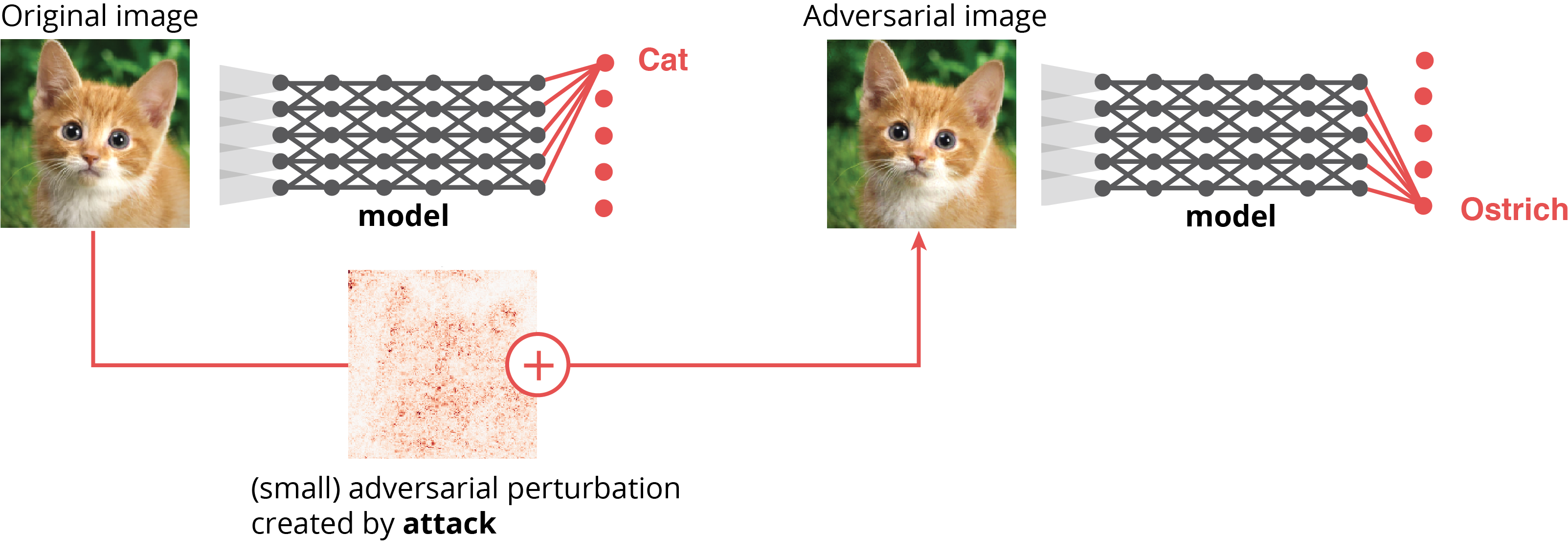}
  \caption{Example of an adversarial image. The original image is classified as a cat. Addition of a carefully designed noise changes the same classifier's output to ostrich, while  the visual change in the image is not significant.}
  \label{fig:adversarial-example}
\end{figure*}

Over the last few years, several studies have suggested that the DNN-based image classifier is highly vulnerable to deception attack \cite{akhtar2018threat, eykholt2017robust}. In fact, with the emergence of internet-of-things (IoT) providing an IP address to all gadgets including cameras, the autonomous vehicles will become more vulnerable to such attacks \cite{chernikova2019self}. Hackers can easily tamper with the pixel values (see Figure~\ref{fig:adversarial-example}) or the image data sent by the camera to the classifier. In a similar way, networked surveillance cameras will also become vulnerable to such malicious attacks.

In order to address the above challenge,  we propose a new  class of   algorithms for adversarial image detection. Our first perturbation-based algorithm PERT performs PCA (Prinicipal Component Analysis) on clean image data set, and detects an adversary by perturbing a test image in the spectral domain along certain carefully chosen coordinates obtained from PCA. Next, its adaptive version APERT  chooses the number of perturbations adaptively in order to minimize the expected number of perturbations subject to constraints on the false alarm and missed detection probabilities.  Numerical results demonstrate the efficacy  of these two algorithms.

\subsection{Related work}
The existing research on adversarial images can be divided into two categories: attack design and attack mitigation.

\subsubsection{\textbf{Attack design}}
\hspace{0.2cm}While there have been numerous attempts to tackle deception attacks in sensor-based remote estimation systems \cite{chattopadhyay2019security, chattopadhyay2018secure, chattopadhyay2018attack}, the problem of design and mitigation of adversarial attack on images to cause misclassification is relatively new. The first paper on adversarial image generation was reported in \cite{szegedy2013intriguing}. Since then, there have been significant research on attack design in this setting. All  these attack schemes can be divided into two categories:
\begin{enumerate}
    \item {\bf White box attack:} Here the attacker knows the architecture, parameters, cost functions, etc of the classifier. Hence, it is easier to design such attacks. Examples of such attacks are given in  
    \cite{goodfellow2014explaining}, \cite{szegedy2013intriguing}, \cite{carlini2017towards}, \cite{madry2017towards}, \cite{papernot2016limitations}, \cite{kurakin2016adversarial}.
    \item {\bf Black box attack:} Here the adversary has access only to the output (e.g., logits or probabilities) of the classifier against a test input. Hence, the attacker has to probe the classifier with many test input images in order to estimate the sensitivity of the output with respect to the input. One black box attack is reported in \cite{brendel2017decision}.
\end{enumerate}

On the other hand, depending on attack goals, the attack schemes can be divided into two  categories:
\begin{enumerate}
\item {\bf Targeted attack:} Such attacks  seek to misclassify a particular class to another pre-defined class. For example, a fruit classifier is made to classify all the apple images as banana. Such attacks are reported in \cite{carlini2018audio} and \cite{brendel2017decision}.
\item {\bf Reliability Attack:} Such attacks only seek to increase the classification error. Such attacks have been reported in \cite{yuan2019adversarial}, \cite{brendel2017decision}, \cite{goodfellow2014explaining}, \cite{madry2017towards}, \cite{szegedy2013intriguing}.
\end{enumerate}
Some popular adversarial attacks are summarized below:

\begin{itemize}
  \item \textbf{L-BFGS Attack  \cite{szegedy2013intriguing}:} 
This white box attack tries to find a perturbation $\bm{\eta}$ to an image $\bm{x}$ such that the perturbed image $\bm{x}'=\bm{x}+\bm{\eta}$ minimizes a cost function $J_{\bm{\theta}}(\bm{x},l)$ (where $\bm{\theta}$ is cost parameter and $l$ is the label) of the classifier, while $\bm{\eta}$ remains within some small set around the origin to ensure small perturbation. A Lagrange multiplier is used to relax the constraint on $\bm{\eta}$, which is found via  line search.
  \item \textbf{Fast Gradient Sign Method (FGSM)  \cite{goodfellow2014explaining}:} Here the perturbation is computed as
  $\bm{\eta} = \epsilon (\nabla_{\bm{x}} J_{\bm{\theta}}(\bm{x},l))$ 
  where $\epsilon$ the magnitude of perturbation. This
  perturbation can be computed via backpropagation.

\textbf{Basic Iterative Method (BIM) \cite{kurakin2016adversarial}:}  This is an iterative variant of FGSM.

  \item \textbf{$\mathbf{Carlini Wagner (L_2) (CW)}$ Attack  \cite{carlini2017towards}:} This is similar to \cite{szegedy2013intriguing} except that: (i) \cite{carlini2017towards} uses a cost function that is different from the classifier's cost function $J_{\bm{\theta}}(\cdot,\cdot)$, and (ii) the optimal Lagrange multiplier is found via binary search.
  \item \textbf{Projected Gradient Descent (PGD) \cite{madry2017towards}:} This involves applying FGSM iteratively and clipping the iterate images to ensure that they remain close to the original image.
  \item \textbf{Jacobian Saliency Map Attack (JSMA) \cite{papernot2016limitations}:} It is a greedy attack algorithm which selects the most important  pixels by calculating Jacobian based saliency map, and  modifies those pixels iteratively.

\item \textbf{Boundary Attack  \cite{brendel2017decision}:}  This is a black box attack which starts from an adversarial point and then performs a random walk along the decision boundary between the adversarial and the non-adversarial regions,   such that the iterate image stays in the adversarial region but the distance between the iterate image and the target image is progressively minimized. This is done via rejection sampling using a suitable proposal distribution, in order  to find progressively smaller adversarial perturbations.
\end{itemize}

\subsubsection{\textbf{Attack mitigation}}
There are two possible approaches for defence against adversarial attack:
\begin{enumerate}
    \item \textbf{Robustness based defense:}  These methods seek  to classify adversarial images correctly, e.g., \cite{xie2019feature},  \cite{papernot2016distillation}.
    \item \textbf{Detection based defense:}  These methods seek  to just distinguish between adversarial and clean images; eg., \cite{feinman2017detecting}, \cite{song2017pixeldefend}.  
    \end{enumerate}
    
  Here we describe some popular attack mitigation schemes. 
 The authors of \cite{xie2019feature} proposed feature denoising  to improve robustness of CNNs against adversarial images. They found that certain architectures were good for robustness  even though they are not sufficient for accuracy improvements. However, when combined with adversarial training, these designs could be more robust.
    The authors of \cite{feinman2017detecting} put forth a Bayesian view of detecting adversarial samples, claiming that the uncertainty associated with adversarial examples is more compared to clean ones. They used a Bayesian neural network to distinguish between adversarial and clean images  on the basis of uncertainty estimation.

The authors of \cite{song2017pixeldefend} trained a PixelCNN network \cite{salimans2017pixelcnn++} to differentiate between clean and adversarial examples. They rejected adversarial samples using p-value based ranking of PixelCNN. This scheme was  able to detect several attacks like $CW(L_2)$, Deepfool, BIM. The paper  \cite{wang2018detecting} observed that there is a significant difference between the percentage of label change due to perturbation in adversarial samples as compared to clean ones. They designed a statistical adversary detection algorithm called nMutant;     inspired by mutation testing from software engineering community. 

The authors of \cite{papernot2016distillation} designed a method called {\em network distillation} to defend DNNs against adversarial examples. 
The original purpose of network distillation was the reduction of size of DNNs by transferring knowledge from a bigger network to a smaller one \cite{ba2014deep}, \cite{hinton2015distilling}. The authors discovered that using high-temperature softmax reduces the model's sensitivity towards small perturbations. This defense was tested on the MNIST and CIFAR-10 data sets. It was observed that network distillation reduces the success rate of JSMA attack \cite{papernot2016limitations} by $0.5 \%$ and 
$5 \%$ respectively. However,   a lot of new attacks have been proposed since then,  which defeat defensive distillation (e.g., \cite{carlini2016defensive}).  The paper  \cite{goodfellow2014explaining} tried training an MNIST classifier with adversarial examples (adversarial retraining approach). A comprehensive analysis  of this method on ImageNet data set found it to be effective against one-step attacks (eg., FGSM), but  ineffective against iterative attacks (e.g., BIM  \cite{kurakin2016adversarial}). After evaluating network distillation with adversarially trained networks  on MNIST and ImageNet,  \cite{tramer2017ensemble} found it to be robust against white box attacks but not against black box ones.

\subsection{Our Contributions}
In this paper, we make the following contributions:
\begin{enumerate}
    \item We propose  a novel detection algorithm PERT for adversarial attack detection. The algorithm performs PCA on clean image data set to obtain a set of orthonormal bases. Projection of a test image along some least significant principal components are randomly perturbed for detecting proximity to a decision boundary, which is used for detection. This combination of PCA and image perturbation in spectral domain, which is motivated by the empirical findings in \cite{hendrycks2016early}, is new  to the literature.\footnote{The paper \cite{liang2017deep} uses PCA but throws away least significant components, thereby removing useful information along those components, possibly leading to high false alarm rate. The paper \cite{carlini2017adversarial} showed that their attack can break simple PCA-based defence, while our  algorithm performs well against the attack of \cite{carlini2017adversarial} as seen later in the numerical results.} 
    \item PERT has low computational complexity; PCA is performed only once off-line.
    \item We also propose an adaptive version of PERT called APERT. The APERT algorithm declares an image to be adversarial by checking whether a specific sequential probability ratio exceeds an upper or a lower threshold. The problem of minimizing the expected number of perturbations per test image, subject to constraints on false alarm and missed detection probabilities, is relaxed via a pair of Lagrange multipliers. The relaxed problem is solved via simultaneous perturbation stochastic approximation (SPSA; see \cite{spall1992multivariate}) to obtain the optimal threshold values, and the optimal Lagrange multipliers are learnt via two-timescale stochastic approximation \cite{borkar2009stochastic} in order to meet the constraints. The use of stochastic approximation and SPSA to optimize the threshold values are new to the signal processing literature to the best of our knowledge. Also, the APERT algorithm has a sound theoretical motivation which is rare in most papers on adversarial image detection.
    \item PERT and APERT are agnostic to attacker and classifier models, which makes  them  attractive to many practical applications.
    \item Numerical results demonstrate high probability of attack detection and small value for false alarm probability under PERT and APERT against a competing algorithm, and reasonably low computational complexity in APERT. 
\end{enumerate}

\subsection{Organization}
The rest of the paper is organized as follows. The PERT algorithm is described in Section~\ref{section:detection-of-single-adversarial-image}. The APERT algorithm is described in Section~\ref{section:sequential-detection}. Numerical exploration of the proposed algorithm is summarized in Section~\ref{section:numerical-results}, followed by the conclusion in Section~\ref{section:conclusion}.

\section{Static perturbation based algorithm}\label{section:detection-of-single-adversarial-image}
\begin{figure}[t!]
  \includegraphics[width=6.5cm, height=4cm]{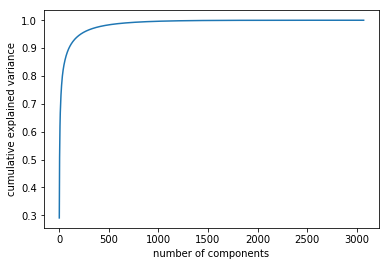}
  \caption{Cumulative explained variance versus components of PCA.}
  \label{fig:PCA-variance-vs-components}
\end{figure}

In this section, we propose an adversarial image detection  algorithm based on random perturbation of an image in the spectral domain; the algorithm is called PERT. This  algorithm is motivated by the two key observations: 
\begin{enumerate}
    \item The authors of \cite{hendrycks2016early}  found that the injected   adversarial noise mainly resides in least significant principal components. Intuitively, this makes sense since injecting noise to the most significant principal components would lead to detection by human eye.  We have applied PCA on CIFAR-10 training data set to learn its principal components sorted by decreasing eigenvalues; the ones with higher eigenvalues are the most significant principal components. CIFAR-10 data set consists of 3072 dimensional images,  applying PCA on the entire data set yields 3072 principal components. The cumulative explained variance ratio as a function of the number of components (in decreasing order of the eigenvalues) is shown in Figure~\ref{fig:PCA-variance-vs-components}; this figure shows that most of the variance is concentrated along the first few principal components. Hence, least significant components do not provide much additional information, and adversarial perturbation of  these components should not change the image significantly.
    \item Several attackers intend  push the image close to the decision boundary  to fool a classifier \cite{brendel2017decision}.  Thus it is possible to detect an adversarial image if we can check whether it is close to a decision boundary or not. Hence, we propose a new scheme for  exploring the neighborhood of a given image in spectral domain.
\end{enumerate}

Hence, our algorithm performs PCA on a training data set, and finds the principal components. When a new test image (potentially adversarial) comes, it projects that image along these principal components, randomly perturbs the projection  along a given number of least significant components, and then obtains another image from this perturbed spectrum. If the classifier yields same label for this new image and the original test image, then it is concluded that the original image is most likely not near a decision boundary and hence not an adversarial; else, an alarm is raised for adversarial attack. In fact, multiple perturbed images can be generated by this process,  and if the label  of the original test image differs with that of at least one perturbed image, an alarm is raised. 
 The intuition behind this is that if an image is adversarial it will lie close to a decision boundary, and perturbation should push it to another region, thus changing the label generated by the classifier.

\begin{algorithm}[h]
\caption{The PERT algorithm}
 \noindent {\bf Training Phase (PCA):} \\
{\em Input: Training image set}\\
{\em Output: Principal components of the data set}\\
\begin{enumerate}
    \item Vectorize  the pixel values of all images.
    \item  Find the sample covariance matrix of these vectors.
    \item Perform singular value decomposition (SVD) of the sample covariance matrix.
    \item Obtain the eigenvectors $\{\bm{\Phi}_1, \bm{\Phi}_2, \cdots, \bm{\Phi}_M\}$ arranged from most significant components to least significant components.
\end{enumerate}

\noindent {\bf Test Phase (Perturbation based attack detection):} 

\SetAlgoLined
\SetKwInOut{Initialization}{Initialization}
\Initialization{Boolean \textit{result} = \textbf{False}}
\KwIn{Input image $\bm{x}$ (vectorized), no. of purturbed image samples to generate $T$, no. of coefficients to perturb $C$}
\KwOut{\textbf{True}, if input is adversarial\\
   \qquad  \qquad  \textbf{False}, if input is not adversarial}
    Get prediction $\bm{y}$ for input image $\bm{x}$ through classifier.\\
  Compute the projections (dot products)  $<\bm{x}, \bm{\Phi}_1>, <\bm{x}, \bm{\Phi}_2>, \cdots, <\bm{x}, \bm{\Phi}_M>$ and vectorize these $M$ values as $\hat{\bm{x}} =[\hat{x}_1,\hat{x}_2,\cdots, \hat{x}_M]$.\\
  
  \For{$i=1$ to $T$}{
  Add realizations of $C$ i.i.d. zero-mean Gaussian random variables to $\hat{x}_{M-C+1},\hat{x}_{M-C+2},\cdots, \hat{x}_M$. This will convert $\hat{\bm{x}}$ to  $\hat{\bm{x}}'$.\\
  Get inverse transform of $\hat{\bm{x}}'$ to get a new image $\bm{x}_i'$.\\
  Get prediction $\bm{y}_i'$ for image $\bm{x}_i'$  through classifier.\\
  \eIf{$\bm{y}$ not equal $\bm{y}_i'$}{
   \textit{result} = \textbf{True}\;
   break\;
   }{
   \textbf{continue}\;
  }
 }
 
\end{algorithm}

{\bf Discussion:} 
PERT has several advantages over most algorithms in the literature:
\begin{enumerate}
    \item PERT is basically a pre-processing algorithm for the test image, and hence it is agnostic to the attacker and classifier models.
    \item The on-line part of PERT involves computing simple dot products and perturbations, which have very low complexity. PCA can be performed once off-line and used for ever.
\end{enumerate}
However, one should remember that PERT perturbs the least significant components randomly, and hence there is no guarantee that a perturbation will be in the right direction to ensure a crossover of the decision boundary. This issue can be resolved by developing more sophisticated perturbation methods using direction search, specifically in case some knowledge of the decision boundaries is available to the detector. Another option is to create many perturbations of a test image, at the expense of more computational complexity. However, in the next section, we will formulate the sequential version of PERT, which will minimize the mean number of image perturbations per image, under  a budget on the missed detection probability and false alarm probability.

\section{Adaptive perturbation based algorithm}\label{section:sequential-detection}
In Section~\ref{section:detection-of-single-adversarial-image}, the PERT algorithm used up to a  constant $T$ number of perturbations of the test image in the spectral domain. However, the major drawback of PERT is that it might be wasteful in terms of computations. If   an adversarial image is very close to the decision boundary, then small number of perturbations might be sufficient for detection. On the other hand, if the adversarial image is far away from a decision boundary, then more perturbations will be required to cross the decision boundary with high probability. Also, the PERT algorithm only checks for a decision  boundary crossover (hard decision), while many DNNs yield a belief probability vector for the class of a test image (soft output); this soft output of DNNs can be used to improve detector performance and reduce its complexity.

In this section, we propose an adaptive version of PERT called  APERT. The APERT algorithm sequentially perturbs the test image in spectral domain. A stopping rule is  used by the pre-processing unit to decide when to stop perturbing a test image and declare a decision (adversarial or non-adversarial); this stopping rule is a two-threshold rule motivated by the sequential probability ratio test (SPRT \cite{poor2013introduction}), on top of the decision boundary crossover checking.  The threshold values are optimized using the theory of stochastic approximation \cite{borkar2009stochastic} and   SPSA \cite{spall1992multivariate}.

\subsection{Mathematical formulation}

Let us denote the random number of perturbations used in any adaptive technique $\mu$ based on random perturbation by $N$, and let the probabilities of false alarm and missed detection of any randomly chosen test image under this technique be denoted by $P_F(\mu)$ and $P_M(\mu)$ respectively. We seek to solve the following constrained problem:
\begin{equation}\label{eqn:constrained-problem}
\min_{\mu} \mathbb{E}_{\mu}(N) \textit{ such that } P_F(\mu) \leq \alpha, P_M(\mu) \leq \beta \tag{CP}
\end{equation}
where $\alpha \in (0,1)$ and $\beta \in (0,1)$ are two constraint values. However, \eqref{eqn:constrained-problem} can be relaxed by using two Lagrange multipliers $\lambda_1 \geq 0, \lambda_2 \geq 0$ to obtain the following unconstrained problem:
\begin{equation}\label{eqn:unconstrained-problem}
\min_{\mu} \bigg(\mathbb{E}_{\mu}(N) + \lambda_1 P_F(\mu) +\lambda_2 P_M(\mu) \bigg) \tag{UP}
\end{equation}
Let the optimal decision rule for \eqref{eqn:unconstrained-problem} under $(\lambda_1, \lambda_2)$ be denoted by $\mu^*(\lambda_1, \lambda_2)$. 
It is well known that, if there exists $\lambda_1^* \geq 0$ and $\lambda_2^* \geq 0$   such that $P_F(\mu^*(\lambda_1^*, \lambda_2^*)) = \alpha, P_M(\mu^*(\lambda_1^*, \lambda_2^*)) = \beta$, then $\mu^*(\lambda_1^*, \lambda_2^*)$ is an optimal solution for \eqref{eqn:constrained-problem} as well. 

Finding out $\mu^*(\lambda_1, \lambda_2)$ for a pair $(\lambda_1, \lambda_2)$ is very challenging. Hence, we focus on the class of SPRT-type algorithms instead. Let us assume that the DNN based classifier generates a probability value against an input image; this probability is the belief of the classifier that the image under consideration is adversarial. Now, suppose that we sequentially perturb an image in the spectral domain as in PERT, and feed these perturbed images one by one to the DNN, which acts as our classifier. Let the DNN return category wise probabilistic distribution of the image in the form of a vector. We use these vectors to determine $q_i$ which indicates the likelihood (not necessarily a probability) of the $i$-th perturbed image being adversarial. Motivated by SPRT, the proposed APERT algorithm checks if  the ratio $\frac{q_1 q_2 \cdots q_k}{|(1-q_1)(1-q_2)\cdots (1-q_k)|}$ crosses an upper threshold $B$ or a lower threshold $A$ after the $k$-th perturbation;  an adversarial image is declared if $\frac{q_1 q_2 \cdots q_k}{|(1-q_1)(1-q_2)\cdots (1-q_k)|}>B$,  a non-adversarial image is declared if $\frac{q_1 q_2 \cdots q_k}{|(1-q_1)(1-q_2)\cdots (1-q_k)|}<A$, and the algorithm continues perturbing the image  if $\frac{q_1 q_2 \cdots q_k}{|(1-q_1)(1-q_2)\cdots (1-q_k)|} \in (A,B)$. In case $k$ exceeds a pre-determined maximum number of perturbations $T$   without any threshold crossing, the image is declared to be non-adversarial.

\begin{algorithm}[h]
\caption{\bf SRT($A,B,C, T, \bm{x}, \{\bm{\Phi_i}\}_{1 \leq i \leq M}, Q)$ algorithm}
\vspace{0.5mm}

\SetAlgoLined
\SetKwInOut{Initialization}{Initialization}
\Initialization{$j=1$, Boolean \textit{result} = \textbf{False}}
\KwIn{Threshold pair $(A,B)$, number  of coefficients to perturb $C$, maximum number  of perturbations $T$, input image $\bm{x}$ (vectorized), orthonormal basis vectors $\{\bm{\Phi_i}\}_{1 \leq i \leq M}$ (typically obtained  from PCA), Switch for Category change detection $Q$ where $Q \in \{0, 1 \}$ }
\KwOut{\textbf{True}, if input image is adversarial\\
   \qquad  \qquad  \textbf{False}, if input image is not adversarial}
    Get category wise probability classification vector $\bm{y}$ for input image $\bm{x}$ through classifier.
    
  Compute the projections (dot products)  $<\bm{x}, \bm{\Phi}_1>, <\bm{x}, \bm{\Phi}_2>, \cdots, <\bm{x}, \bm{\Phi}_M>$ and vectorize these $M$ values as $\hat{\bm{x}} =[\hat{x}_1,\hat{x}_2,\cdots, \hat{x}_M]$.\\
 
  \While{$j  \leq T$ and $\textit{result} = False$}{
  Add realizations of $C$ i.i.d. zero-mean Gaussian random variables to $\hat{x}_{M-C+1},\hat{x}_{M-C+2},\cdots, \hat{x}_M$. This will convert $\hat{\bm{x}}$ to  $\hat{\bm{x}}'$.
  Get inverse transform of $\hat{\bm{x}}'$ to get a new image $\bm{x}_j'$.
  
  Get category wise probability classification vector $\bm{y}_j'$ for input image $\bm{x}_j'$ through classifier.
  
  Get $q_j$ by taking $\frac{\|\bm{y} - \bm{y}_j'\|_p}{K}$, where $K$ is the dimension of row vector $\bm{y}$ and $1 \leq p $ 
  
  \uIf{Predicted category changed in perturbed image and $Q = 1$}{
    \textit{result} = \textbf{True}\;
    break\;
  }
  \uElseIf{$\frac{q_1 q_2 .. q_j}{|(1-q_1)....(1-q_j)|} < A$}
   {\textit{result} = \textbf{False}\;
   break\;}
  \uElseIf{$\frac{q_1 q_2 .. q_j}{|(1-q_1)....(1-q_j)|} > B$}
   {\textit{result} = \textbf{True}\;
   break\;}
  \Else{
   $j = j + 1$
   \textbf{continue}\;
  
  }
}
\Return \textit{result}
\end{algorithm}

Clearly, for  given $(\lambda_1, \lambda_2)$, the algorithm needs to compute the optimal threshold values $A^*(\lambda_1, \lambda_2)$ and $B^*(\lambda_1, \lambda_2)$ to minimize the cost in \eqref{eqn:unconstrained-problem}. Also, $\lambda_1^*$ and  $\lambda_2^*$ need to be computed to meet the constraints in \eqref{eqn:constrained-problem} with equality. APERT uses two-timescale stochastic approximation and SPSA for updating the Lagrange multipliers and the threshold values in the training phase, learns the optimal parameter values, and uses these parameter values in the test phase.  

\subsection{The SRT algorithm for image classification}
Here we describe an SPRT based algorithm called sequential ration test or SRT for classifying an image $\bm{x}$. The algorithm takes $A,B, C, T, \bm{x}$, the PCA eigenvectors $\{\bm{\Phi_i}\}_{1 \leq i \leq M}$, and a binary variable $Q \in \{0,1\}$ as input, and classifies $\bm{x}$ as adversarial or non-adversarial. This algorithm is used as one important component of the APERT algorithm described later. SRT blends ideas from PERT and the standard SPRT algorithm. However, as seen in the pseudocode of SRT, we use a quantity $q_j$ in the  threshold testing where $q_j \in (0,1)$ cannot be interpreted as  a probability. Instead, $q_j$ is the normalized value of the $p$-norm of the difference between outputs $\bm{y}$ and $\bm{y}_j'$ of the DNN against inputs $\bm{x}$ and its $j$-th   perturbation $\bm{x}_j'$. The binary variable $Q$ is used as a switch; if $Q=1$ and if the belief probability vectors $\bm{y}$ and $\bm{y}_j'$ lead to two different predicted categories, then SRT directly declares $\bm{x}$ to be adversarial. It has been observed numerically  that this results in a better adversarial image detection probability, and hence any test image in the proposed APERT scheme later is classified via SRT with $Q=1$.

\subsection{The  APERT algorithm}
\subsubsection{The training phases} 
The APERT algorithm, designed for \eqref{eqn:constrained-problem},  consists of two training phases and a testing phase. The first training phase   simply runs the PCA algorithm. The second training phase basically runs stochastic approximation iterations to find $A^*, B^*, \lambda_1^*, \lambda_2^*$ so that the false alarm and missed detection probability constraints are satisfied with equality.

The second training phase of APERT   requires three non-negative sequences   $a(t)$, $\delta(t)$ and $d(t)$ are chosen such that: (i)    $\sum_{t=0}^{\infty}a(t)$ = $\sum_{t=0}^{\infty}d(t)$ = $\infty$, (ii) 
    $\sum_{t=0}^{\infty}a^2(t) < \infty$, $\sum_{t=0}^{\infty}d^2(t) < \infty$, (iii) 
    $\lim_{t\to\infty}\delta(t) = 0$, (iv) 
    $\sum_{t=0}^{\infty}\frac{a^2(t)}{\delta^2(t)} < \infty$, (v) $\lim_{t\to\infty}\frac{d(t)}{a(t)} = 0$. The first two conditions are standard requirements for stochastic approximation. The third and fourth conditions are required for convergence of SPSA, and the fifth condition maintains the necessary timescale separation explained later.
    
    The APERT algorithm also requires $\theta$ which is the percentage of adversarial images among all image samples used in the training phase II. It also maintains  two iterates $n_{clean}(t)$ and $n_{adv}(t)$ to represent the number of clean and images encountered up to the $t$-th training image; i.e., $n_{clean}(t)+n_{adv}(t)=t$.  
    
    Steps $5-13$ of APERT correspond to SPSA which is basically a stochastic gradient descent scheme with noisy estimate of  gradient, used for minimizing the objective of \eqref{eqn:unconstrained-problem} over $A$ and $B$  for current $(\lambda_1(t), \lambda_2(t))$ iterates. SPSA allows us to compute a noisy gradient of the objective of \eqref{eqn:constrained-problem} by randomly and simultaneously perturbing $(A(t), B(t))$ in two opposite directions and obtaining  the noisy estimate of gradient from the difference in the objective function evaluated at these two perturbed values; this allows us to avoid coordinate-wise perturbation in gradient estimation. In has to be noted that, the cost to be optimized by SPSA has to be obtained from SRT.    The $A(t)$ and $B(t)$ iterates are projected onto non-overlapping compact intervals $[A_{min}, A_{max}]$ and $[B_{min}, B_{max}]$ (with $A_{max}<B_{min}$) to ensure boundedness.
    
    Steps~$14-15$ are used to find $\lambda_1^*$ and $\lambda_2^*$ via stochastic approximation in a slower timescale. In has to be noted that, since $\lim_{t \rightarrow \infty}\frac{d(t)}{a(t)}=0$, we have a two-timescale stochastic approximation \cite{borkar2009stochastic} where the Lagrange multipliers are updated in a slower timescale and the threshold values are updated via SPSA in a faster timescale. The faster timescale iterates view the slower timescale iterates as quasi-static, while the slower-timescale iterates view the faster timescale iterates as almost equilibriated; as if, the slower timescale iterates vary in an outer loop and the faster timescale iterates vary in an inner loop. It has to be noted that, though standard two-timescale stochastic approximation theory guarantees some convergence under suitable conditions \cite{borkar2009stochastic}, here we cannot provide any convergence guarantee of the iterates due to the lack of established  statistical properties of the images. It is also noted that, $\lambda_1(t)$ and $\lambda_2(t)$ are updated at different time instants; this corresponds to asynchronous stochastic approximation \cite{borkar2009stochastic}. The $\lambda_1(t)$ and $\lambda_2(t)$ iterates are projected onto $[0,\infty)$ to ensure non-negativity.  Intuitively, if a false alarm is observed, the cost of false alarm, $\lambda_1(t)$ is increased. Similarly, if a missed detection is observed, then the cost of missed detection, $\lambda_2(t)$, is increased, else it is decreased. Ideally, the goal is to reach to a pair $(\lambda_1^*,\lambda_2^*)$ so that the constraints in \eqref{eqn:constrained-problem} are met with equality, through we do not have any formal convergence proof.
    
    \subsubsection{Testing phase}
    The testing phase just uses SRT with $Q=1$ for any test image. Since $Q=1$, a test image bypasses the threshold testing and is declared adversarial, in case the random perturbation results in predicted category change of the test image. It has been numerically observed (see Section~\ref{section:numerical-results}) that this results in a small increase in false alarm rate but a high increase in adversarial image detection rate compared to $Q=0$. However, one has the liberty to avoid this and only use the threshold test in SRT by setting $Q=0$. Alternatively, one can set a slightly smaller value of $\alpha$ in APERT with $Q=1$ in order to compensate for the increase in false alarm.

\begin{algorithm}
\caption{\bf The APERT algorithm}
 \vspace{0.5mm}
 
{ \small
 \noindent {\bf Training Phase I (PCA):} \\
{\em Input: Training image set}\\
{\em Output: Principal components of the data set}\\
\begin{enumerate}
    \item Vectorize  the pixel values of all images.
    \item  Find the sample covariance matrix of these vectors.
    \item Perform singular value decomposition (SVD) of the sample covariance matrix.
    \item Obtain the eigenvectors $\{\bm{\Phi}_1, \bm{\Phi}_2, \cdots, \bm{\Phi}_M\}$ arranged from most significant components to least significant components.
\end{enumerate}

\noindent{\bf Training Phase II (Determining A and B)}

\SetAlgoLined
\SetKwInOut{Initialization}{Initialization}
\Initialization{$n_{clean}(0) = n_{adv}(0) = 0$,  non-negative $0<A(0)< B(0), \lambda_1(0), \lambda_2(0)$}
\KwIn{$\theta$, training image set with each image in vectorized form, number of training images $n_{max}$, no. of coefficients to perturb $C$,  $\alpha$, $\beta$,  sequences  $\{a(t), \delta(t), d(t)\}_{t \geq 0}$, maximum number of perturbations  $N_{max}$, range of accepted values of the thresholds $[A_{min}, A_{max}]$, $[B_{min}, B_{max}]$ such that $0 < A_{min}<A_{max} < B_{min}<B_{max}$}
\KwOut{Final Values of $(A^*, B^*)$ (and also $\lambda_1^*$ and $\lambda_2^*$ which are not used in the test phase)}

\For{$t\gets0$ \KwTo $n_{max}-1$}{
    Randomly generate $b_1(t) \in  \{-1, 1\}$ and $b_2(t) \in \{-1, 1\}$ with probability $0.5$
    
    Compute $A'(t) = A(t) + \delta(t)b_1(t)$ and $B'(t) = B(t) + \delta(t)b_2(t)$, similarly $A''(t)= A(t) - \delta(t)b_1(t)$ and 
    $B''(t)= B(t) - \delta(t)b_2(t)$.
    
    Randomly pick the training image $\bm{x_t}$ (can be adversarial with probability $\theta/100$)
    
    If image is actually adversarial then $n_{adv}(t) = n_{adv}(t-1) + 1$, If image is clean then $n_{clean}(t) = n_{clean}(t-1) + 1$
    
    Define $\mathbb{I}\{miss\}$ and $\mathbb{I}\{falsealarm\}$ as an indicator of missed detection and false alarm  respectively by SRT algorithm and $\mathbb{I}\{cleanimage\}$ as indicator of a non adversarial image
    
    Compute cost $c'(t) =N'(t)+\lambda_1(t) \mathbb{I}\{falsealarm\}+\lambda_2(t) \mathbb{I}\{miss\} $ by using $SRT(A'(t), B'(t), C, T,\bm{x}_t, \{\bm{\Phi_i}\}_{1 \leq i \leq M}, 0)$ 
    
    Compute cost $c''(t) =N''(t)+\lambda_1(t) \mathbb{I}\{falsealarm\}+\lambda_2(t) \mathbb{I}\{miss\} $ by using $SRT(A''(t), B''(t), C, T,\bm{x}_t, \{\bm{\Phi_i}\}_{1 \leq i \leq M}, 0)$ 
    
    Update $A(t+1) = [A(t) - a(t) \times \frac{c'(t) - c''(t)}{2b_1(t)\delta(t)}]$ and then project $A(t+1)$ on $[A_{min}, A_{max}]$  
    
    Update $B(t+1) = [B(t) - a(t) \times \frac{c'(t) - c''(t)}{2b_2(t)\delta(t)}]$ and then project $B(t+1)$ on $[B_{min}, B_{max}]$
    
    
    Again Determine $\mathbb{I}\{falsealarm\}$ and $\mathbb{I}\{miss\}$ from SRT$(A(t+1), B(t+1), C, T,\bm{x_t}, \{\bm{\Phi_i}\}_{1 \leq i \leq M}, 0)$
    
    Update $\lambda_1(t+1) = [\lambda_1(t) +\mathbb{I}\{clean image\} d(n_{clean}(t)) \times (\mathbb{I}\{falsealarm\} - \alpha)]_0^{\infty}$, $\lambda_2(t+1) = [\lambda_2(t) + \mathbb{I}\{adversarial image\} d(n_{adv}(t)) \times (\mathbb{I}\{miss\} - \beta)]_0^{\infty}$

     } 
     \Return $(A^* \doteq A(n_{max}), B^* \doteq B(n_{max}))$


\noindent{\bf Testing Phase:} 

\SetAlgoLined
\SetKwInOut{Initialization}{Initialization}
\Initialization{Boolean \textit{result} = \textbf{False}}
\KwIn{Input image $\bm{x}$ (vectorized), maximum number of perturbed image samples to generate $T$, no. of coefficients to perturb $C$, lower threshold $A$, upper threshold $B$}
\KwOut{\textbf{True}, if input image is adversarial\\
   \qquad  \qquad  \textbf{False}, if input image is not adversarial}
   
   \Return SRT$(A, B, C, T, \bm{x}, \{\bm{\Phi_i}\}_{1 \leq i \leq M}, 1)$
   
}
\end{algorithm}

\section{Experiments}\label{section:numerical-results}
\subsection{Performance of PERT}
We  evaluated our proposed algorithm on CIFAR-10 data set and the classifier of \cite{madry2017towards}  implemented in a challenge to explore adversarial robustness of neural networks (see \cite{MadryLabCifar10}).\footnote{Codes for our numerical experiments are available in \cite{PCA_detection} and \cite{SPRT_detection}.  We used Foolbox library \cite{rauber2017foolbox}  for generating adversarial images.   PCA was performed using Scikit-learn \cite{scikit-learn} library in Python; this library allows us to customize the computational complexity and accuracy in PCA.} Each image in CIFAR-10 has $32 \times 32$ pixels, where each pixel has three channels: red, green, blue. Hence, PCA provides $M=32 \times 32 \times 3=3072$ orthonormal basis vectors. CIFAR-10 has $60000$ images, out of which $50000$ images were used for PCA based training and rest of the images were used for evaluating  the performance of the algorithm.

Table~\ref{table:performance-versus-number-of-samples}  shows  the variation of detection probability (percentage of detected adversarial images) for adversarial images generated using various attacks, for number of components $C=1000$ and various values for maximum possible number of  samples $T$ (number of perturbations for a given image). Due to huge computational requirement in generating adversarial images via black box attack, we have considered only four white box attacks. It is evident that the attack detection probability (percentage) increases with $T$; this is intuitive since larger $T$ results in a higher probability of decision boundary crossover if an adversarial image is perturbed. The second column of Table~\ref{table:performance-versus-number-of-samples} denotes the percentage of clean images that were declared to be adversarial by our algorithm, i.e., it contains the false alarm probabilities which also increase with $T$. However, we observe that our pre-processing algorithm achieves very low false alarm probability and high attack detection probability under these four popular white box attacks. This conclusion is further reinforced in   Table~\ref{table:performance-versus-number-of-components}, which shows the variation in detection performance with varying $C$, for $T=10$ and $T=20$. It is to be noted that the detection probability under the detection algorithm of \cite{wang2018detecting} are $56.7 \%$ and $72.9 \%$ for $CW(L_2)$ and FGSM attacks; clearly our detection algorithm outperforms \cite{wang2018detecting} while having low computation.    The last column of       Table~\ref{table:performance-versus-number-of-components} suggests that there is an optimal value of $C$, since perturbation along more principal components may increase the decision boundary crossover probability but at the same time can  modify the information along some most significant components as well.

\begin{table}[htbp]
\caption{Detection and false alarm performance of PERT algorithm for various values of $T$.}
\begin{center}
\begin{tabular}{|c|c|c|c|c|c|}
\hline
\textbf{No. of} & \multicolumn{5}{|c|}{\textbf{Percentage Detection (\%)}} \\
\cline{2-6} 
\textbf{Samples $T$}& \textbf{Clean$^{\mathrm{a}}$}& \textbf{\textit{FGSM}}& \textbf{\textit{L-BFGS}}& \textbf{\textit{PGD}}& \textbf{\textit{CW(L\textsubscript{2})}} \\
\hline
05 & 1.2 & 50.02 & 89.16 & 55.03 & 96.47\\
\hline
10 & 1.5 & 63.53 & 92.50 & 65.08 & 98.23\\
\hline
15 & 1.7 & 69.41 & 93.33 & 67.45 & 99.41\\
\hline
20 & 1.9 & 73.53 & 95.03 & 71.01 & 99.41\\
\hline
25 & 1.9 & 75.29 & 95.03 & 75.14 & 100.00\\
\hline
\multicolumn{4}{l}{$^{\mathrm{a}}$Clean images that are detected as adversarial}
\end{tabular}
\label{table:performance-versus-number-of-samples}
\end{center}
\end{table}

\begin{table}[htbp]
\caption{Detection and false alarm performance of PERT algorithm for various values of $C$.}
\begin{center}
\begin{tabular}{|c|c|c|c|c|c|}
\hline
\textbf{No. of} & \multicolumn{5}{|c|}{\textbf{Percentage Detection (\%)}} \\
\cline{2-6} 
\textbf{Coefficients $C$}& \textbf{Clean$^{\mathrm{a}}$}& \textbf{\textit{FGSM}}& \textbf{\textit{L-BFGS}}& \textbf{\textit{PGD}}& \textbf{\textit{CW(L\textsubscript{2})}} \\
\hline
\multicolumn{6}{|c|}{\textbf{No. of Samples($T$): 10}} \\
\cline{1-6} 

0500 & 1.20 & 58.23 & 90.83 & 57.40 & 95.90\\
\hline
1000 & 1.50 & 69.41 & 93.33 & 60.95 & 95.45\\
\hline
1500 & 2.10 & 64.11 & 91.67 & 61.53 & 95.00\\
\hline
\multicolumn{6}{|c|}{\textbf{No. of Samples($T$): 20}} \\
\cline{1-6} 
0500 & 1.20 & 68.23 & 93.33 & 68.05 & 95.90\\
\hline
1000 & 1.90 & 74.11 & 94.16 & 70.41 & 95.90\\
\hline
1500 & 2.50 & 71.18 & 95.00 & 71.00 & 95.00\\
\hline
\multicolumn{4}{l}{$^{\mathrm{a}}$Clean images that are detected as adversarial}
\end{tabular}
\label{table:performance-versus-number-of-components}
\end{center}
\end{table}


\subsection{Performance of APERT}

For APERT, we initialize $A(0) = 10^{-10}, B(0) = 0.5, \lambda_1(0) = \lambda_2(0) = 10$, and choose step sizes  $a(t) =O (\frac{1}{t^{0.7}}), d(t) = O(\frac{1}{t}), \delta(t) = O(\frac{1}{t^{0.1}})$.   The Foolbox library was used to craft adversarial examples. The classification neural network is  taken from \cite{MadryLabCifar10}

$2$-norm is used to obtain the $q_i$ values since it was observed that $2$-norm outperforms $1$-norm. In the training process, $50 \%$ of the training images were clean and $50\%$ images were adversarial.

Though there is no theoretical convergence guarantee for APERT, we have numerically observed convergence of $A(t)$, $B(t)$, $\lambda_1(t)$ and $\lambda_2(t)$

\subsubsection{Computational complexity of PERT and APERT}
We note that, a major source of computational complexity in PERT and APERT is perturbing an image and passing it through a classifier. In Table~\ref{table:APERT vs PERT Mean Sample Peformance} and Table~\ref{table:APERT vs PERT Mean Sample Peformance Q = 0}, we numerically compare the mean number of perturbations required for PERT and APERT under $Q=1$ and $Q=0$ respectively. The classification neural network was taken from \cite{MadryLabCifar10}.

Table~\ref{table:APERT vs PERT Mean Sample Peformance} and Table~\ref{table:APERT vs PERT Mean Sample Peformance Q = 0} show that the APERT algorithm requires much less perturbations compared to PERT for almost similar detection performance, for various attack algorithms and various test images that result in false alarm, adversarial image detection, missed detection and (correctly) clean image detection. It is also noted that, for the images resulting in missed detection and clean image detection, PERT has to exhaust all $T=25$ perturbation options before stopping. As a result, the mean number of perturbations in APERT becomes significantly smaller than PERT; see Table~\ref{table:APERT vs PERT Mean Sample Peformance over full Dataset}. The key reason behind smaller number of perturbations in APERT is the fact that APERT uses a doubly-threshold stopping rule motivated by the popular SPRT algorithm in detection theory. It is also observed that APERT with $Q=1$ in the testing phase has slightly lower computaional complexity than APERT with $Q=0$, since APERT with $Q=1$ has an additional flexibility of stopping the perturbation if there is a change in predicted category.

\begin{table}[h]
\caption{Mean number of Samples generated for PERT and APERT algorithm.Here PERT and APERT's parameter were set in a way that they bring their false alarm performance closest to each other for $Q = 1$ in testing phase for APERT}
\begin{center}
\begin{adjustbox}{width=0.5\textwidth}
\begin{tabular}{|c|c|c|c|c|c|c|c|c|}
\hline
\textbf{Attack} & \multicolumn{8}{|c|}{\textbf{Mean Number of Samples Generated}} \\
\cline{2-9} 
\textbf{Type}& \multicolumn{2}{|c|}{\textbf{False}}& \multicolumn{2}{|c|}{\textbf{Detected}} & \multicolumn{2}{|c|}{\textbf{Missed}} & \multicolumn{2}{|c|}{\textbf{Detected}}\\

\textbf{}& \multicolumn{2}{|c|}{\textbf{Alarm}}& \multicolumn{2}{|c|}{\textbf{Adversarial}} & \multicolumn{2}{|c|}{\textbf{Detection}} & \multicolumn{2}{|c|}{\textbf{Clean}}\\
\hline
 & \textbf{PERT} & \textbf{APERT} & \textbf{PERT} & \textbf{APERT} & \textbf{PERT} & \textbf{APERT} & \textbf{PERT} & \textbf{APERT} \\
\hline
$CW(L_2)$ & 9.76 & 1.17 & 1.19 & 1.02 & 25 & 4.09 & 25 & 2.37\\
\hline
LBFGS & 11.86 & 1.42 & 1.87 & 1.07 & 25 & 4.97 & 25 & 3.41\\
\hline
FGSM & 1.68 & 1.08 & 4.97 & 1.07 & 25 & 5.08 & 25 & 2.97\\
\hline
PGD & 14.12 & 1.15 & 4.87 & 1.41 & 25 & 5.47 & 25 & 3.03\\
\hline
\textbf{Attack} & \multicolumn{8}{|c|}{\textbf{Corresponding Detection performance \%}} \\
\cline{2-9} 
\textbf{Type}& \multicolumn{2}{|c|}{\textbf{False}}& \multicolumn{2}{|c|}{\textbf{Detected}} & \multicolumn{2}{|c|}{\textbf{Missed}} & \multicolumn{2}{|c|}{\textbf{Detected}}\\

\textbf{}& \multicolumn{2}{|c|}{\textbf{Alarm Probability}}& \multicolumn{2}{|c|}{\textbf{Adversarial Probability}} & \multicolumn{2}{|c|}{\textbf{Detection Probability}} & \multicolumn{2}{|c|}{\textbf{Clean Probability}}\\
\hline
 & \textbf{PERT} & \textbf{APERT} & \textbf{PERT} & \textbf{APERT} & \textbf{PERT} & \textbf{APERT} & \textbf{PERT} & \textbf{APERT} \\
\hline
$CW(L_2)$ & 4.56 & 5.12 & 97.10 & 98.10 & 2.90 & 1.90 & 95.44 & 94.88\\
\hline
LBFGS & 4.85 & 5.24 & 96.3 & 94.35 & 3.7 & 5.65 & 95.15 & 94.76\\
\hline
FGSM & 5.41 & 5.88 & 79.31 & 87.64 & 20.69 & 12.36 & 94.59 & 94.12\\
\hline
PGD & 4.01 & 4.51 & 83.99 & 84.45 & 16.01 & 15.55 & 95.99 & 95.49\\
\hline

\end{tabular}
\end{adjustbox}
\label{table:APERT vs PERT Mean Sample Peformance}
\end{center}
\end{table}

\begin{table}[h]
\caption{Mean number of Samples generated for PERT and APERT algorithm.Here APERT's parameters were set in a way that they bring the false alarm performance of APERT closest to corresponding PERT's false alarm performance in Table~\ref{table:APERT vs PERT Mean Sample Peformance} for $Q = 0$ in testing phase for APERT}
\begin{center}
\begin{adjustbox}{width=0.5\textwidth}
\begin{tabular}{|c|c|c|c|c|c|c|c|c|}
\hline
\textbf{Attack} & \multicolumn{8}{|c|}{\textbf{Mean Number of Samples Generated}} \\
\cline{2-9} 
\textbf{Type}& \multicolumn{2}{|c|}{\textbf{False}}& \multicolumn{2}{|c|}{\textbf{Detected}} & \multicolumn{2}{|c|}{\textbf{Missed}} & \multicolumn{2}{|c|}{\textbf{Detected}}\\

\textbf{}& \multicolumn{2}{|c|}{\textbf{Alarm}}& \multicolumn{2}{|c|}{\textbf{Adversarial}} & \multicolumn{2}{|c|}{\textbf{Detection}} & \multicolumn{2}{|c|}{\textbf{Clean}}\\
\hline
 & \textbf{PERT} & \textbf{APERT} & \textbf{PERT} & \textbf{APERT} & \textbf{PERT} & \textbf{APERT} & \textbf{PERT} & \textbf{APERT} \\
\hline
$CW(L_2)$ & 9.76 & 1.0 & 1.19 & 1.0 & 25 & 6.11 & 25 & 2.96\\
\hline
LBFGS & 11.86 & 1.02 & 1.87 & 1.0 & 25 & 6.07 & 25 & 3.28\\
\hline
FGSM & 1.68 & 1.05 & 4.97 & 1.02 & 25 & 6.11 & 25 & 3.10\\
\hline
PGD & 14.12 & 1.07 & 4.87 & 1.21 & 25 & 6.037 & 25 & 3.05\\
\hline
\textbf{Attack} & \multicolumn{8}{|c|}{\textbf{Corresponding Detection performance \%}} \\
\cline{2-9} 
\textbf{Type}& \multicolumn{2}{|c|}{\textbf{False}}& \multicolumn{2}{|c|}{\textbf{Detected}} & \multicolumn{2}{|c|}{\textbf{Missed}} & \multicolumn{2}{|c|}{\textbf{Detected}}\\

\textbf{}& \multicolumn{2}{|c|}{\textbf{Alarm Probability}}& \multicolumn{2}{|c|}{\textbf{Adversarial Probability}} & \multicolumn{2}{|c|}{\textbf{Detection Probability}} & \multicolumn{2}{|c|}{\textbf{Clean Probability}}\\
\hline
 & \textbf{PERT} & \textbf{APERT} & \textbf{PERT} & \textbf{APERT} & \textbf{PERT} & \textbf{APERT} & \textbf{PERT} & \textbf{APERT} \\
\hline
$CW(L_2)$ & 4.56 & 5.45 & 97.10 & 84.09 & 2.90 & 15.91 & 95.44 & 94.55\\
\hline
LBFGS & 4.85 & 5.97 & 96.3 & 80.57 & 3.7 & 19.43 & 95.15 & 94.03\\
\hline
FGSM & 5.41 & 6.47 & 79.31 & 65.29 & 20.69 & 34.71 & 94.59 & 93.53\\
\hline
PGD & 4.01 & 5.12 & 83.99 & 65.49 & 16.01 & 34.51 & 95.99 & 94.88\\
\hline

\end{tabular}
\end{adjustbox}
\label{table:APERT vs PERT Mean Sample Peformance Q = 0}
\end{center}
\end{table}

\begin{table}[h]
\caption{Mean number of Samples generated for T = 25 and C = 1000 for APERT and PERT algorithm over a Dataset with $50\%$ Aversarial images and $50\%$ clean images }
\begin{center}
\begin{tabular}{|c|c|c|c|}
\hline
 & \multicolumn{3}{|c|}{\textbf{Mean Number of Samples Generated}} \\
\cline{2-4} 
\hline
\textbf{Attack} & \textbf{PERT} & \multicolumn{2}{|c|}{\textbf{APERT}} \\
\cline{3-4}
\textbf{} &\textbf{} & \textbf{Q = 1} & \textbf{Q = 0} \\
 \hline
$CW(L_2)$ & 13.3 & 1.85 & 1.92\\
\hline
LBFGS & 13.5 & 2.19 & 2.19\\
\hline
FGSM & 15.5 & 2.14 & 2.56\\
\hline
PGD & 14.48 & 2.20 & 2.57\\
\hline

\end{tabular}
\label{table:APERT vs PERT Mean Sample Peformance over full Dataset}
\end{center}
\end{table}

\begin{figure*}[t]
  \includegraphics{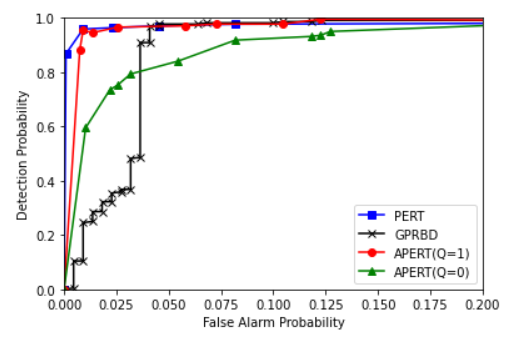}
  \includegraphics{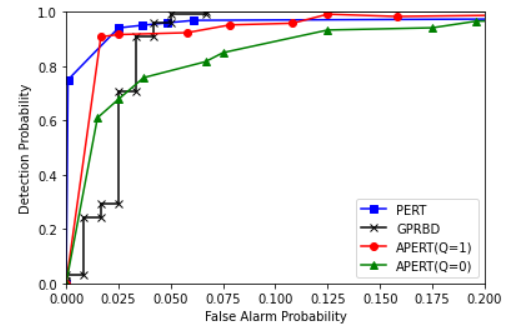}
  \includegraphics{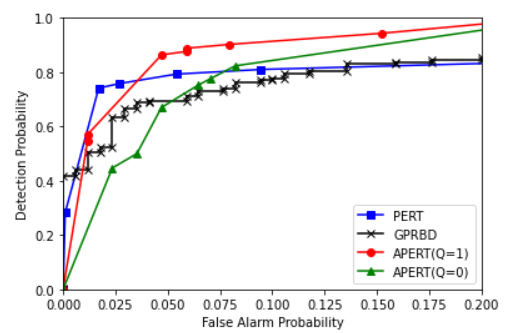}
  \includegraphics{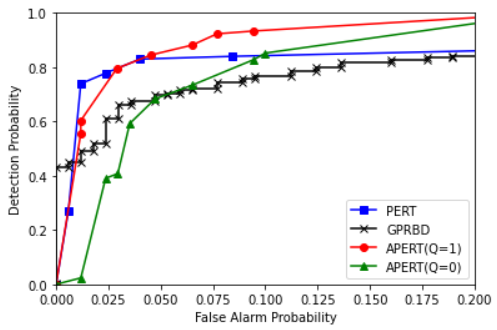}
  \caption{ROC plot comparison of PERT, APERT with Q = 0, APERT with Q = 1 and GPRBD  detection algorithms for various attack schemes. Top left: CW attack. Top right: LBFGS attack. Bottom left: FGSM attack. Bottom right: PGD attack.}
  \label{fig:ROC-comparisons}
\end{figure*}

We also implemented a Gaussian process regression based detector (GPRBD) from \cite{lee2019adversarial} (not sequential in nature) which uses the neural network classifier of \cite{MadryLabCifar10}, tested it against our adversarial examples, and compared its runtime against that of PERT and APERT equipped with the neural network classifier of \cite{MadryLabCifar10}. These experiments were run under the same colab runtime environment, in a single session. The runtine specifications are- CPU Model name: Intel(R) Xeon(R) CPU @ 2.30GHz, Socket(s): 1, Core(s) per socket: 1, Thread(s) per core: 2, L3 cache: 46080K, CPU MHz: 2300.000, RAM available: 12.4 GB, Disk Space Available: 71 GB. Table~\ref{table:Runtime of GPY} shows that, APERT has significantly  smaller runtime than PERT as expected, and slightly larger runtime than GPRBD. Also, APERT with $Q=1$ has smaller runtime than $Q=0$.

\begin{table}[h]
\caption{Performance of our implementation of Gaussian Process Regression based detector(GPRBD) vs our APERT algorithm for T = 25, C = 1000 }
\begin{center}
\begin{tabular}{|c|c|c|c|c|}
\hline
\textbf{Attack} & \multicolumn{4}{|c|}{\textbf{Average Time Taken per Image (seconds)}} \\
\hline
 & \textbf{GPRBD} & \multicolumn{2}{|c|}{\textbf{APERT}} & \textbf{PERT}\\
\hline
 & & Q = 1 & Q = 0 & \\
 \hline
$CW(L_2)$ & 0.2829 & 0.6074 & 0.6398 & 4.1257\\
\hline
LBFGS & 0.2560 & 0.6982 & 0.7059 & 4.7895\\
\hline
FGSM & 0.2728 & 0.6372 & 0.7801 & 4.6421\\
\hline
PGD & 0.2694 & 0.6475 & 0.7789 & 4.4216\\
\hline

\end{tabular}
\label{table:Runtime of GPY}
\end{center}
\end{table}

\subsubsection{Performance of PERT and APERT}
In Figure~\ref{fig:ROC-comparisons}, we compare the ROC (receiver operating characteristic) plots of PERT, APERT and GPRBD algorithms, all implemented with the same neural network classifier of \cite{MadryLabCifar10}. The Gaussian model used for GPRBD was implemented using \cite{gpy2014} with the Kernel parameters set as follows: input dimensions = 10, variance = 1 and length scale = 0.01 as in \cite{lee2019adversarial}. The Gaussian model  parameter optimization was done using LBFGS with max iterations = 1000. It is obvious from Figure~\ref{fig:ROC-comparisons} that, for the same false alarm probability, APERT has higher or almost same attack detection rate compared to PERT. Also, APERT and PERT significantly outperform GPRBD. Hence, APERT yields a good compromise between ROC performance and computational complexity. It is also observed that APERT with $Q=1$ always has a better ROC curve than APERT with $Q=0$ in the testing phase.

Table~\ref{table:APERT Detection variation over C with T = 10 and 20} and Table~\ref{table:APERT Detection variation over C with T = 10 and 20 for Q = 0} show that the false alarm probability and attack detection probability of APERT increases with $C$ for a fixed $T$, for both $Q=1$ and $Q=0$. As $C$ increases, more least significant components are perturbed in the spectral domain, resulting in a higher probability of decision boundary crossover.

\begin{table}[h]
\caption{Variation in Performance of APERT with values of C using second norm for T = 10 and T = 20 for Q = 1 in testing phase}
\begin{center}

\begin{adjustbox}{width=0.5\textwidth}
\begin{tabular}{|c|c|c|c|c|c|c|}
\hline
\textbf{Attack} & \multicolumn{6}{|c|}{\textbf{Percentage Detection(\%)}} \\
\cline{2-7} 
\textbf{Type}& \multicolumn{2}{|c|}{\textbf{C = 500}}& \multicolumn{2}{|c|}{\textbf{C = 1000}} & \multicolumn{2}{|c|}{\textbf{C = 1500}} \\

\hline
 & \textbf{False} & \textbf{Detection} & \textbf{False} & \textbf{Detection} & \textbf{False} & \textbf{Detection} \\
 & \textbf{Alarm} & \textbf{Probability} & \textbf{Alarm} & \textbf{Probability} & \textbf{Alarm} & \textbf{Probability} \\
\hline
\multicolumn{7}{|c|}{\textbf{No. of Samples($T$): 10}} \\
\cline{1-7} 
$CW(L_2)$ & 5.0 & 96.81 & 8.18 & 98.18 & 11.36 & 99.09\\
\hline
LBFGS & 4.16 & 94.16 & 10.0 & 93.33 & 12.5 & 97.5\\
\hline
FGSM & 2.94 & 62.35 & 4.70 & 80.0 & 5.88 & 94.70\\
\hline
PGD & 2.36 & 63.9 & 5.91 & 77.54 & 7.10 & 88.16\\
\hline
\multicolumn{7}{|c|}{\textbf{No. of Samples($T$): 20}} \\
\cline{1-7} 
$CW(L_2)$ & 1.01 & 95.46 & 5.45 & 96.81 & 6.36 & 98.18\\
\hline
LBFGS & 2.5 & 90.83 & 9.16 & 94.16 & 15.0 & 96.66\\
\hline
FGSM & 2.94 & 57.64 & 4.70 & 79.41 & 7.64 & 90.58\\
\hline
PGD & 1.77 & 60.9 & 4.14 & 79.88 & 9.46 & 88.757\\
\hline
\end{tabular}
\end{adjustbox}
\label{table:APERT Detection variation over C with T = 10 and 20}
\end{center}

\end{table}

\begin{table}[h]
\caption{Variation in Performance of APERT with values of C using second norm for T = 10 and T = 20, for Q = 0 in testing phase}
\begin{center}

\begin{adjustbox}{width=0.5\textwidth}
\begin{tabular}{|c|c|c|c|c|c|c|}
\hline
\textbf{Attack} & \multicolumn{6}{|c|}{\textbf{Percentage Detection(\%)}} \\
\cline{2-7} 
\textbf{Type}& \multicolumn{2}{|c|}{\textbf{C = 500}}& \multicolumn{2}{|c|}{\textbf{C = 1000}} & \multicolumn{2}{|c|}{\textbf{C = 1500}} \\

\hline
 & \textbf{False} & \textbf{Detection} & \textbf{False} & \textbf{Detection} & \textbf{False} & \textbf{Detection} \\
 & \textbf{Alarm} & \textbf{Probability} & \textbf{Alarm} & \textbf{Probability} & \textbf{Alarm} & \textbf{Probability} \\
\hline
\multicolumn{7}{|c|}{\textbf{No. of Samples($T$): 10}} \\
\cline{1-7} 
$CW(L_2)$ & 3.18 & 89.09 & 8.18 & 91.81 & 12.72 & 93.63\\
\hline
LBFGS & 6.66 & 81.66 & 15.0 & 85.83 & 19.16 & 96.66\\
\hline
FGSM & 2.35 & 50.0 & 7.05 & 67.65 & 8.23 & 85.88\\
\hline
PGD & 2.36 & 39.05 & 6.5 & 68.63 & 9.46 & 82.84\\
\hline
\multicolumn{7}{|c|}{\textbf{No. of Samples($T$): 20}} \\
\cline{1-7} 
$CW(L_2)$ & 5.45 & 84.09 & 8.18 & 91.81 & 11.81 & 93.18\\
\hline
LBFGS & 7.5 & 85.0 & 12.5 & 93.33 & 17.5 & 94.166\\
\hline
FGSM & 3.529 & 44.70 & 6.47 & 65.29 & 7.05 & 83.53\\
\hline
PGD & 2.95 & 40.82 & 7.10 & 67.45 & 10.0 & 85.21\\
\hline
\end{tabular}
\end{adjustbox}
\label{table:APERT Detection variation over C with T = 10 and 20 for Q = 0}
\end{center}

\end{table}

\section{Conclusion}\label{section:conclusion}
In this paper, we have proposed two novel pre-processing schemes for detection of adversarial images, via a combination of PCA-based spectral decomposition, random perturbation, SPSA and two-timescale stochastic approximation. The proposed  schemes have reasonably low computational complexity and are agnostic to attacker and classifier models.  Numerical results on detection and false alarm probabilities  demonstrate the efficacy of the proposed algorithms, despite having low computational complexity. We will extend this  work for detection of black box attacks in our future research endeavour.

{\small
\bibliographystyle{unsrt}
\bibliography{sample-base}
}

\end{document}